\documentclass{article}
\usepackage{graphicx}
\usepackage{amssymb}
\usepackage{amsmath}
\usepackage{booktabs}
\usepackage{url}
\usepackage{xspace}
\usepackage{nicefrac}
\usepackage{capt-of}
\usepackage{setspace}
\usepackage{subfigure}

\usepackage{pstricks-add}
\usepackage{pstricks}
\usepackage{pst-tree}

\usepackage{algorithmic}
\usepackage[algo2e,ruled, vlined, linesnumbered, commentsnumbered]{algorithm2e}

\usepackage{multirow}

\begin{document}

\title{A model of macro-evolution as a branching process based on innovations}

\author{Stephanie Keller-Schmidt and Konstantin Klemm\\
{\small Bioinformatics, Institute for Computer Science,}\\
{\small Leipzig University, H\"{a}rtelstrasse 16-18, 04107 Leipzig, Germany}}

\maketitle

\begin{abstract}
We introduce a model for the evolution of species triggered by generation of
novel features and exhaustive combination with other available traits. Under the
assumption that innovations are rare, we obtain a bursty  branching process of
speciations. Analysis of the trees representing the branching history reveals
structures qualitatively different from those of random processes. For a tree
with $n$ leaves generated by the introduced model, the average distance of leaves from root scales as $(\log n)^2$
to be compared to $\log n$ for random branching.  The mean values and standard
deviations for the tree shape indices depth (Sackin index) and imbalance
(Colless index) of the model are compatible with those of real
phylogenetic trees from databases. Earlier models, such as the Aldous' branching
(AB) model, show a larger deviation from data with respect to the shape
indices.
\end{abstract}

%%%%%%%%%%%%%%%%%%%%%%%%%%%%%%%%%%%%%%%%%%%%%%%%%%%%%%%%%%%%%%%%%%%%%%%%%%%%
\section{Introduction} \label{sec:intro}
%%%%%%%%%%%%%%%%%%%%%%%%%%%%%%%%%%%%%%%%%%%%%%%%%%%%%%%%%%%%%%%%%%%%%%%%%%%%

Since the seminal work by Darwin \cite{Darwin:1859}, the evolution of biological
species has been recognized as a complex dynamics involving broad distributions
of temporal and spatial scales as well as stochastic effects, giving rise to
so-called frozen accidents. There is vast exchange and overlap of concepts and
methods between the theory of evolution and the foundations of complex systems
such as fitness landscapes \cite{Wright:1932,Gavrilets:2004,Klemm:2012} and
neutral networks \cite{Kimura:1983}, the evolution of cooperation
\cite{Axelrod:1984} and self-organized criticality \cite{Bak:1996} to name but a
few.

A striking feature of biological macroevolution is its burstiness. The temporal
distribution of speciation and extinction events is highly inhomogeneous in time
\cite{Sepkoski:1993}. As described by the theory of punctuated equilibrium
\cite{Gould:1993}, a connection between punctuated equilibrium in evolution and
the theory of self-organized criticality \cite{Bak:1996} is established through
the model by Bak and Sneppen \cite{Bak:1993,Sneppen:1995}. Ecology, i.e.\ the
system of trophic interactions and other dependencies between species'
fitnesses, is driven to a critical state. Then minimal perturbations cause
relaxation cascades of broadly distributed sizes.

Rather than through ecological interaction across possibly all species, bursty
diversification may also be due to {\em adaptive radiation} as a
rapid multiplication of species in one lineage after a triggering event. About
200 million years ago, a novel chewing system with dedicated molar teeth
evolved in the lineage of mammals, allowing it to rapidly diversify into species
using vastly distinct types of nutrition \cite{Ungar:2010}. There are many more
examples where a single {\em innovation} triggers adaptive
radiation such as the tetrapod limb morphology caused by a binary shift in bone arrangement
\cite{Thomson1992Macroevolution} and the homeothermy as a key innovation by the group of
mammals \cite{HeardHauser1995keyInnovations,Leim1990innovationHomeothermy}.
Environmental conditions a species has not encountered previously,
e.g.\ when entering a geographical area with unoccupied ecological niches, may
also be the source of adaptive radiation. The diversity of finch species on
Galapagos islands is the famous example first studied by Darwin. Spontaneous
phenotypic or genetic innovations and those caused by the pressure to adapt to a
change in environment are treated on the same footing for the modeling purposes
in this contribution. Though being a central concept in the theory of
evolution, the term innovation has not been ascribed a unique definition so far
\cite{Pigliucci:2008}.

Here we study a branching process to mimic the evolution of species driven
by innovations. The process involves a separation of time scales.
Rare innovation events trigger rapid cascades of diversification where a 
feature combines with previously existing features. We call this newly defined
branching process {\em innovation model}.

How can the validity of models of this kind be assessed? The evolutionary
history of species is captured by phylogenetic trees. These are binary
trees where leaves represent extant species, alive today, and inner nodes stand
for ancestral species from which the extant species have descended.
By comparing the shapes of these trees
\cite{Sackin1972phenogram,Herrada2008universalScaling,Campos:2004,Stich:2009},
in particular their degree of imbalance
\cite{Colless1982phylogenetics,Mckenzie2000DistributionCherries},
with trees generated by different evolutionary mechanisms
\cite{Aldous2001FromYuleToToday,Blum2006whichRandomProcess,HernandezGarcia2010scaling}, a selection of
realistic models is possible.

%%%%%%%%%%%%%%%%%%%%%%%%%%%%%%%%%%%%%%%%%%%%%%%%%%%%%%%%%%%%%%%%%%%%%%%%%%%%
\section{Stochastic models of macroevolution}
%%%%%%%%%%%%%%%%%%%%%%%%%%%%%%%%%%%%%%%%%%%%%%%%%%%%%%%%%%%%%%%%%%%%%%%%%%%%

We consider models of macroevolution within the following formal framework.
At each point in time $t$, there is a set of species $S(t)$. Evolution proceeds
as follows. A species $s \in S(t)$ is chosen according to a probability
distribution $\pi(s,t)$ on $S(t)$. Speciation of $s$ means replacing $s$ by two
new species $s^\prime$ and $s^{\prime\prime}$ such that
\begin{equation}
S(t+1) = S(t) \setminus \{s\} \cup \{s^\prime, s^{\prime\prime} \}
\end{equation}
is the set of species at time $t+1$. The initial condition (at $t=1$) is
a single species. Therefore discrete time $t$ and number of species $n$ are
identical, $n=|S(t)|=t$.

\subsection{Trees}

\begin{figure}
\centerline{\includegraphics{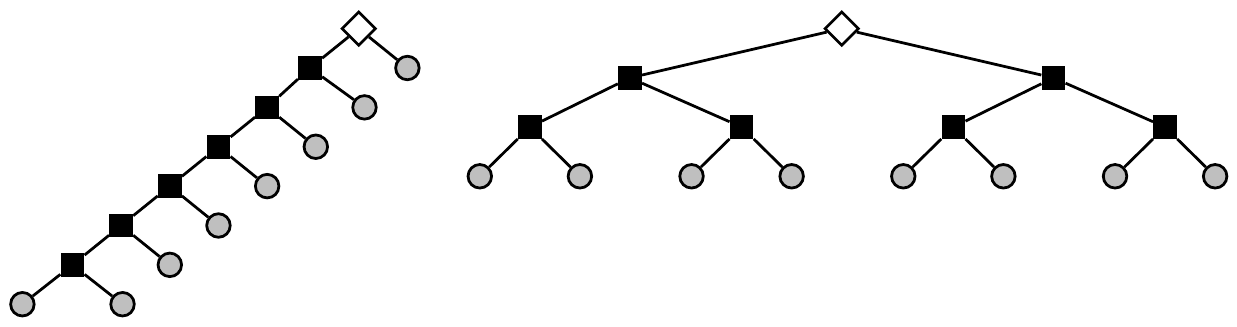}}
\caption{Comparison of tree shapes. Each tree of size eight consists of a root (white diamond), a set of inner nodes (black squares) and a set of leaves (gray circles). The left tree is totally imbalanced, also called comb tree,
with depth $d=35/8 =4.375$ and Colless index $c=21/21=1$~.
The right tree is a complete binary tree with depth $d=24/8=3$ and Colless index $c=0/21=0$~.}
\label{fig:balance}
\end{figure}

The evolutionary history of organisms is represented by a phylogenetic tree. For
the purpose of this contribution, a phyologenetic tree is a rooted strict binary
tree $T$: a tree with exactly one node (the root) with degree two or zero, all
other nodes having degree three (inner node) or one (leaf node), cf. illustrations in Figure \ref{fig:balance}. For such a tree
$T$ with root $w$, a subtree $T^\prime$ is obtained as the component not
containing $w$ after cutting an edge $\{i,j\}$ of $T$. $T^\prime$ is again a
rooted strict binary tree. Since this contribution focuses on tree shape,
all edges have unit length. The distance between nodes 
$i$ and $j$ on a tree $T$ is the number of edges contained in the unique
path between $i$ and $j$.

From the evolutionary dynamics, an evolving phylogenetic tree $T(t)$ is obtained
as follows. At each time step $t$, the leaves of $T(t)$ are the species $S(t)$.
When $s$ undergoes specation, two new leaves $s^\prime$ and $s^{\prime\prime}$
attach to a leaf $s$. After this event, $s$ is an inner node and no longer a
leaf of the tree. In this way, each model of speciation dynamics also defines a
model for the growth of a binary tree by iterative splitting of leaves.

\subsection{Yule model}

In the simplest case, the probability of choosing a species is uniform
at each time step, $\pi(s,t) = 1/t$. This is the Yule model or ERM model.
It serves as a null model of evolution. 

The model corresponds to a particularly simple probability distribution on the
set of generated trees. For a tree with $n \ge 2$ leaves generated by the Yule
model and $i \in \{1,2\dots,n-1\}$, let $p_\text{ERM}(i|n)$ be the probability
that exactly $i$ leaves are in the left subtree of the root. Then
$p_\text{ERM}(i|n) = 1/(n-1)$. This is shown inductively
as follows. Obtaining exactly $i$ leaves
at step $n$, either they were already present at the previous step and the
speciation took place in the right subtree, or the number increased from $i-1$
to $i$ by speciation in the left subtree. Addition of these products of
probabilities for the two cases yields
\begin{equation}
p_\text{ERM}(i|n) =
\frac{n-1-i}{n-1} p_\text{ERM}(i|n-1) +
\frac{i-1}{n-1} p_\text{ERM}(i-1|n-1) ~.
\end{equation}
With the induction hypothesis $p_\text{ERM}(j|n-1)=1/(n-2)$ for all $j$,
we obtain
\begin{equation}
\label{eq:perm}
p_\text{ERM}(i|n) = \frac{(n-1-i)+(i-1)}{(n-1)(n-2)} = \frac{1}{n-1}~.
\end{equation}
The induction starts with $p_\text{ERM}(1|2)=1$ which holds because a tree
with two leaves has one leaf each in the left and in the right subtree.
Thus the uniform selection of species turns into a uniform distribution
on the number of nodes in the left or right subtree. Note that the same
distribution applies to each subtree of an ERM tree. Therefore $p_\text{ERM}$
fully describes the statistical ensemble of ERM trees. The probability
of obtaining a particular tree is the product of $p_\text{ERM}$ terms taken
over all subtrees. This becomes particularly relevant for modifications of the
model taking $p$ non-uniform, see the following subsection.

\subsection{Aldous' branching (AB) model}

The class of beta-splitting models defines a distribution of trees by the
probability
\begin{equation} \label{eq:beta}
p_{\beta}(i|n)=\frac{1}{\alpha_{\beta}(n)}\frac{\Gamma(\beta+l+1)\Gamma(\beta+n-l+1)}{\Gamma(l+1)\Gamma(n-l+1)}
\end{equation}
with appropriate normalization factor $\alpha_{\beta}(n)$. Analogous to $p_\text{ERM}$
of the previous subsection, $p_\beta(i|n)$ is the probability that a
tree has $i$ out of its $n$ leaves in the left subtree. In order to build
a tree with $n$ leaves, one first decides according to $p_\beta(i|n)$
to have $i$ leaves in the left and $n-i$ leaves in the right subtree.
Then the same rule is applied to both subtrees with the determined number
of leaves. The recursion into deeper subtrees naturally stops when a subtree
is decided to have one leaf. 

The parameter $\beta\in [-2;+\infty[$
in Equation~(\ref{eq:beta}) tunes the expected imbalance.
By increasing $\beta$, equitable splits with $i \approx n/2$ become
more probable. The probability distribution of trees from the Yule model
is recovered by taking $\beta=0$. The case $\beta=-1.5$ is called
Proportional to Distinguishable Arrangements (PDA). It produces a uniform
distribution of all ordered (left-right labeled) trees of a given size $n$
\cite{Rosen1978biogeography,Pinelis2003evolutinarymodels,SteelMcKenzie2001propertiesOfPhyloTrees,CottonPage2006ShapeGeneFamilyPhylogenies}.

Another interesting case is Aldous' branching (AB) model
\cite{ALDOUSBetaSplitting1996,Aldous2001FromYuleToToday}
obtained for $\beta=-1$, where Equation~\ref{eq:beta}
reads
\begin{equation}
p_{-1}(i|n) \propto \frac{1}{i(n-i)}~.
\end{equation}
Blum and Fran{\c c}ois have found that $\beta=-1.0$  is the maximum-likelihood
choice of $\beta$ over a large set of phylogenetic trees
\cite{Blum2006whichRandomProcess}. Therefore we use it as a standard of
comparison. The AB model does not have an interpretation in
terms of macroevolution, as noted by Blum and Fran{\c c}ois
\cite{Blum2006whichRandomProcess}. In particular, it is unknown if
its probability distribution of trees can be obtained by stochastic processes 
of iterated speciation as introduced at the beginning of this section.

\subsection{Activity model}

In the activity model \cite{HernandezGarcia2010scaling}, the set 
of species $S(t)$ is partitioned into a set of active species $S_A(t)$ and
a set of inactive species $S_I(t)$. At each time step, a species $s \in S_A(t)$
is drawn uniformly if $S_A(t)$ is non-empty. Otherwise $s \in S_I(t)$ is
drawn uniformly. The two new species
$s^\prime$ and $s^{\prime\prime}$ independently enter the active set $S_A(t+1)$
with probability $p$. The activation probability $p$ is a parameter of the model.
For $p=0.5$ a critical branching process is obtained. Otherwise the model is
similar to the Yule model. A variation of the activity model has been 
introduced by Herrada et al.\ \cite{Herrada:2011} in the context of
protein family trees.

\subsection{Age-dependent speciation}

In the \emph{age model} \cite{KellerSchmidtTugrul2010AgeModel},
the probability of speciation is inversely proportional to the
age of a species. At each time, a species $s \in S(t)$ is
drawn with probability
\begin{equation}
\pi_s(t) \propto \tau_s(t)^{-1}
\end{equation}
normalized properly. The age $\tau_s$ is the number of time steps
passed since creation of species $s$.

\subsection{Innovation model} \label{sec:innov_def}

\begin{algorithm2e}
%	\KwIn{$N$ \ldots\ amount of nodes which simulated tree $T$ should have;\\
%	\hspace{3.5em}$S$ \ldots\ set of all species $s$;\\
%	\hspace{3.5em}$F(t)$ \ldots\ set of all features existing at time $t$;\\
%	}
%	\KwOut{$T$ of size $N$}
 	\BlankLine
 	set $t=1$, $F(0)=\emptyset$, $S(0) = \{\emptyset\}$;\\
 	\While(\tcp*[f]{$N$ as final size of simulated tree}){$|S(t)| < N$}{
	\eIf{$S(t) \setminus \{ s\setminus \{\phi\} : s \in S(t), \phi \in F(t)\} \neq \emptyset$}{
 		\tcp{loss event}
     	draw $\phi \in F(t)$ uniformly;\\
     	draw $s \in S(t)$ uniformly;\\
     	\If{$s \setminus \{\phi\} \notin S(t)$}{
 	 		$S(t+1) = S(t) \cup \{s \setminus \{\phi\}\}$;\\
       		        $F(t+1) = F(t)$;\\
		increment $t$;\\
     	}
  	}{
 	  	\tcp{innovation event}
   		draw $s \in S(t)$ uniformly;\\ 
     	set  $\phi = 1 + \max (F(t) \cup \{0\})$; \\
     	set $S(t+1) = S(t) \cup \{ s \cup \{\phi\}\};$ \\ 
     	set $F(t+1) = F(t) \cup \{\phi\}$;\\
     	increment $t$;\\
  	}
 	}%end while 	
\caption{Pseudocode for the innovation model}
\label{algo:innovmodel}
\end{algorithm2e}

\begin{figure}
\centerline{\includegraphics[width=\textwidth]{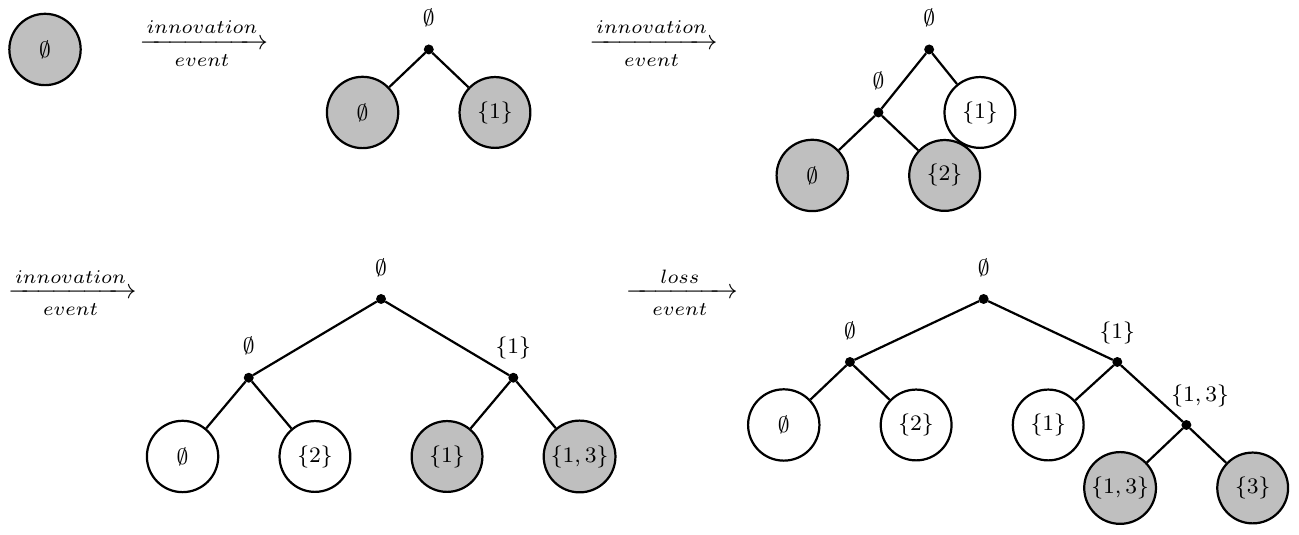}}
\caption{\label{fig:innov_example}
A tree of five leaves generated by the innovation model. The
root node labeled with the empty feature set $\emptyset$ speciates by
an innovation event adding the feature 1 to the feature set. This results
in the species $\emptyset$ and $\{1\}$~. Innovation events are
performed, generating features until a loss event is possible. The
loss event generates the species $\{3\}$ by removing
the feature $1$ from $\{1,3\}$.}
\end{figure}

In the {\em innovation model}, each species $s$ is defined as a finite set of
features $s \subseteq \mathbb{N}$. Features are taken as integer numbers in
order to have an infinite supply of symbols. We denote by $F(t)$ the set of
all features existing at time $t$, that is $F(t)=\bigcup_{s \in S(t)} s$.
Each speciation occurs as one of two possible events. 

An {\em innovation} is the addition of a new feature
$\phi \in \mathbb{N} \setminus F(t)$ not yet contained in any
species at the given time $t$. One of the resulting species
carries the new feature, $s^\prime = s \cup \{\phi\}$. The other
species has the same features as the ancestral one, $s^{\prime\prime}=s$.

A {\em loss} event generates a new species by the
disappearance of a feature. A feature $\phi$ is drawn from $F(t)$ uniformly.
The loss event is performed only if $s \setminus \{\phi\} \notin S(t)$
such that elimination of $\phi$ from $s$ actually generates a new species.
In this case, the resulting species are the one having suffered the loss,
$s^\prime = s \setminus \{\phi\}$ and the species $s^{\prime\prime}=s$
remaining unaltered. Otherwise, $\phi$ is not present in $s$ or its loss
would lead to another already existing species, so nothing happens.

We assume that creation of novel features is significantly less abundant than speciation
by losses. This separation of time scales is implemented by the rule that
an innovation event is only possible when no more losses can be performed.
In order to facilitate further studies with the model, we provide a pseudocode
description in Algorithm~\ref{algo:innovmodel}. Figure~\ref{fig:innov_example}
shows an example of the dynamics.

%%%%%%%%%%%%%%%%%%%%%%%%%%%%%%%%%%%%%%%%%%%%%%%%%%%%%%%%%%%%%%%%%%%%%%%%%%%%
\section{Comparison of simulated and empirical data sets}
%%%%%%%%%%%%%%%%%%%%%%%%%%%%%%%%%%%%%%%%%%%%%%%%%%%%%%%%%%%%%%%%%%%%%%%%%%%%

\begin{figure}
\centering
\includegraphics[width=\textwidth]{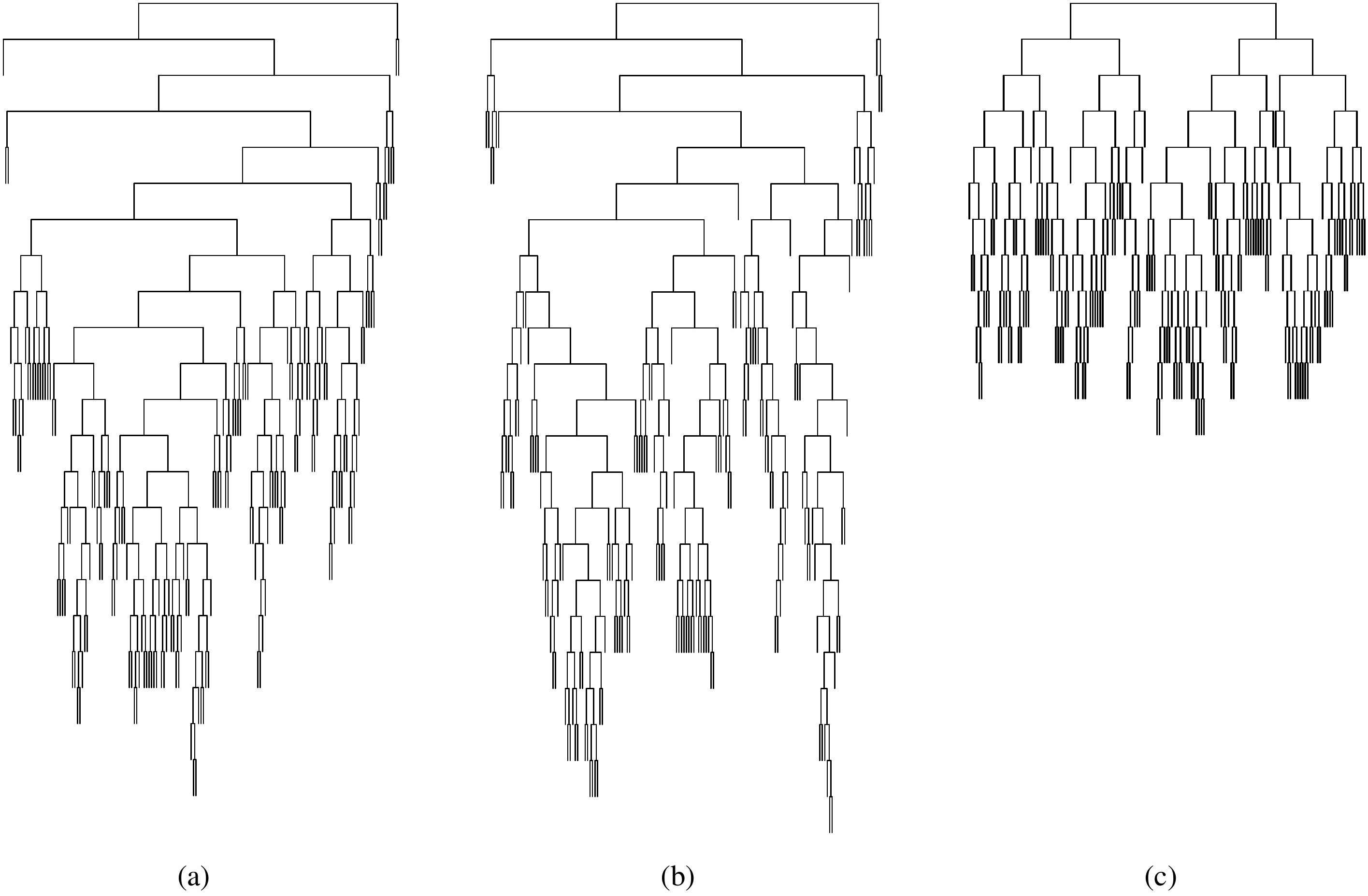}
\caption{\label{fig:treeplots}
Empirical and simulated trees. The depicted phylogenetic tree in (a) is from the database TreeBASE (Matrix ID M2957, relationships in rosids based on mitochondrial
matR sequences), (b) is a tree created as a realization of the innovation model and
(c) a tree from the ERM (Yule) model. Each of the trees has 161 leaves.} 
\end{figure}

Now let us compare the tree shapes obtained by the models with those of
evolutionary trees in databases. The TreeBASE \cite{Sanderson1994treebase}
database contains phylogenetic information about the evolution of species
whereas the database PANDIT \cite{Whelan2006Pandit} contains phylogenetic trees
representing protein domains. Analysing the properties with reference to the
tree shape of both data sets and applying a comparative study with statistical
data sets of different models one can conclude how well a growth model
constructs ``real'' trees.

Comparison by simple inspection of trees from real data and models may already
reveal substantial shape differences. Figure~\ref{fig:treeplots} shows
an example. The trees in panels (a) and (b) are less compact than that of
panel (c) of Figure~\ref{fig:treeplots}. 

For an objective and quantitative comparison of trees, we use the following two
measures of tree shape. Compactness is best described by the distance $d_i$ of a
leaf $i$ from root being small. The {\em depth} (or Sackin index)
\cite{Sackin1972phenogram} is the average distance of leaves from root,
\begin{equation}
d =\frac{\sum_{i=1}^{n} d_i}{n}~.
\end{equation}

The {\em Colless index} measures the average
imbalance of a tree \cite{Colless1982phylogenetics}.
The imbalance at an {\em inner node} $j$ of the tree
is the absolute difference $c_j = |l_j-r_j|$ of leaves in the left and right
subtree rooted at $j$. Then the average of imbalances
\begin{equation}
c= \frac{2}{(n-1)(n-2)} \sum_{j=1}^{n-1} c_j
\end{equation}
with appropriate normalization is the Colless index $c$ of the tree.
The index $j$ runs over all $n-1$ inner nodes including the root itself.
We find $c=0$ for a totally balanced tree and $c=1$ for a comb tree,
see also Figure~\ref{fig:balance}.

\begin{figure}
\centering
\includegraphics[width=\textwidth]{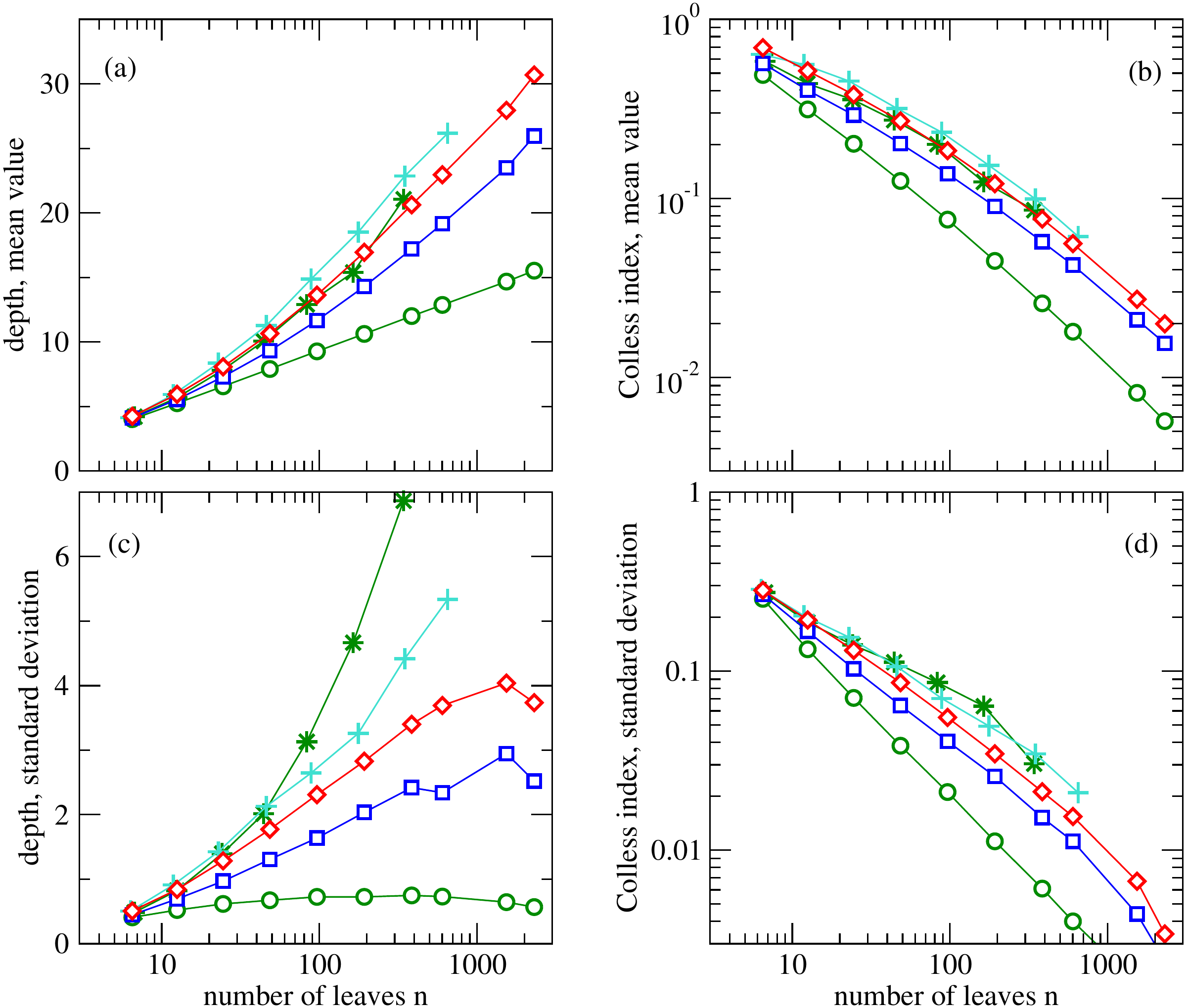}
\caption{\label{fig:comparison}
Comparison of size-dependent summary statistics for models and real trees.
Symbols distinguish the ERM model ($\circ$), the AB model ($\Box$) and the
innovation model ($\diamond$) and the data sets TreeBASE ($\ast$) and PANDIT
($+$). The data sets were preprocessed by solving monotomies and
polytomies randomly as well as removing the outgroups as proposed by
\cite{Blum2006whichRandomProcess}. The mean values of depth, 
and Colless index, panels (a) and (b) are binned logarithmically as a function
of tree size $n$. The same procedure is applied to the standard deviations,
panels (c) and (d).
The analysed TreeBASE data set has been downloaded from http://www.treebase.org
on June, 2007 containing 5,087 trees of size 5 to 535~ after preprocessing. The
PANDIT data set has been downloaded from http://www.ebi.ac.uk/goldman-srv/pandit on May 2008 and includes 36,136 preprocessed trees of size 5 to 2,562~. The simulated data set comprises for each model (AB model, ERM model and innovation model) 1,000
trees for each tree size from 5 to 535 and 10 trees for each tree size from 536
to 2,562. 
} 
\end{figure}

\begin{figure}
\centering
\includegraphics[width=\textwidth]{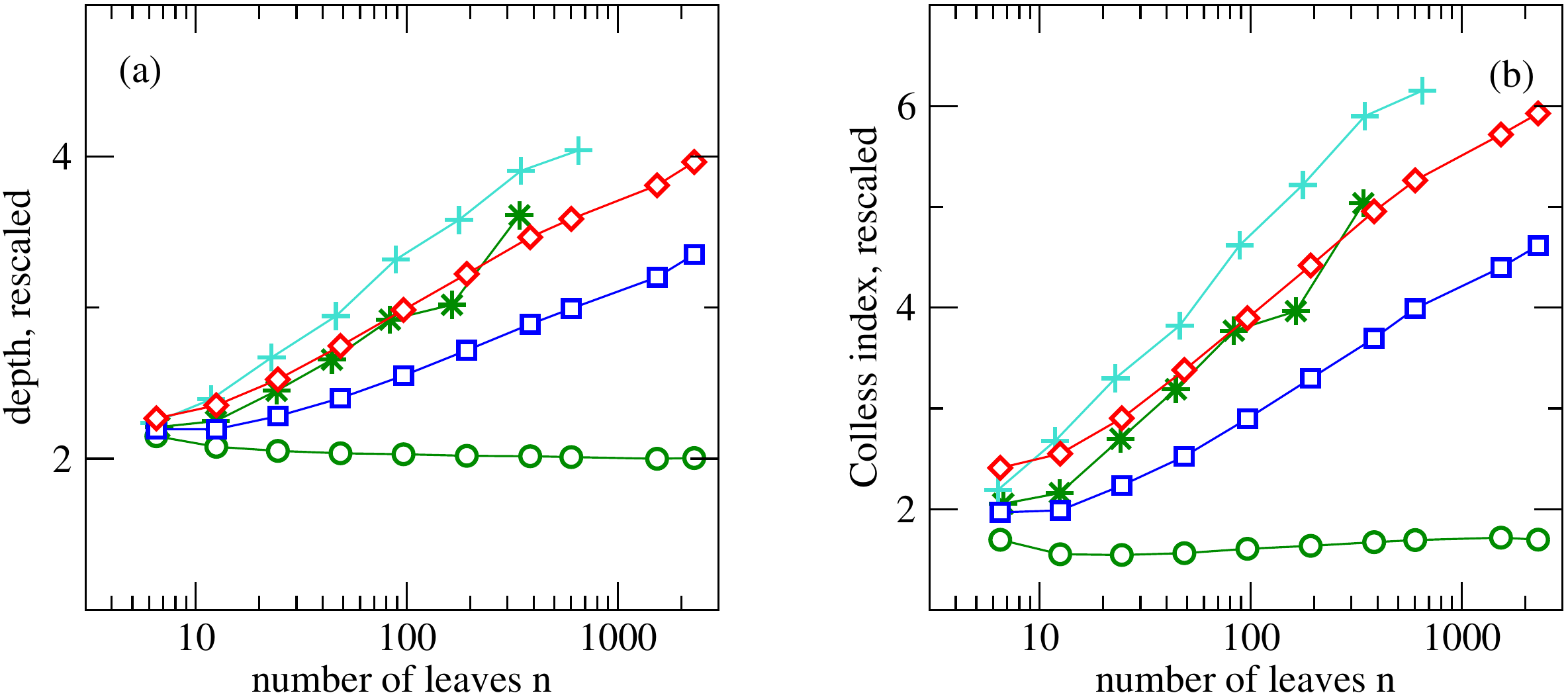}
\caption{\label{fig:rescaled}
The same values of depth and Colless index as in Figure~\ref{fig:comparison}
(a,b) with an $n$-dependent rescaling. (a)
Average depth divided by $\ln n$. (b) Average Colless index divided by
$n^{-1}\ln n$. These factors are chosen such that the rescaled values for
the ERM model asymptotically approach a constant. See refererence 
\cite{BlumFrancoisJanson2007MeanVarianceTwoStatistics} for the scaling
of the indices of the ERM model.}
\end{figure}

Ensemble mean values and standard deviations of these indices are
shown in Figure \ref{fig:comparison}. Comparing the results of three models
(ERM, AB and innovation) to those of trees from two databases, the
least discrepancy is obtained between the innovation model and the
trees from TreeBASE, representing macroevolution. In Figure~\ref{fig:rescaled},
the averages of the two indices are shown after rescaling to facilitate
the comparison. Of all models,
the values of the innovation model are also best matching those
of PANDIT.

%%%%%%%%%%%%%%%%%%%%%%%%%%%%%%%%%%%%%%%%%%%%%%%%%%%%%%%%%%%%%%%%%%%%%%%%%%%%
\section{Depth scaling in the innovation model}
%%%%%%%%%%%%%%%%%%%%%%%%%%%%%%%%%%%%%%%%%%%%%%%%%%%%%%%%%%%%%%%%%%%%%%%%%%%%

\subsection{Subtree generated by an innovation}

\begin{figure}
\centerline{\includegraphics[width=\textwidth]{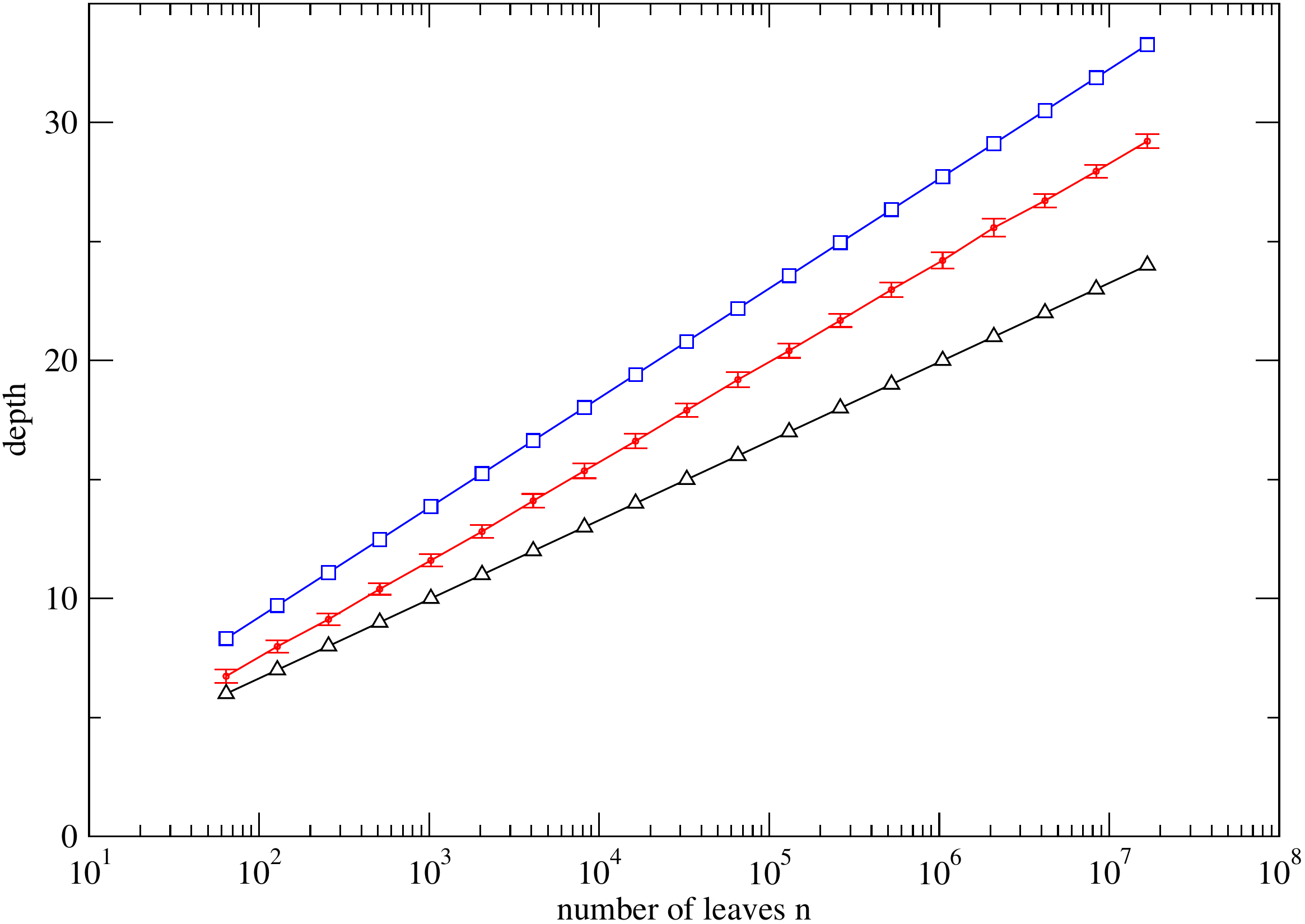}}
\caption{ \label{fig:depth_scaling_loss}
Average depth in dependence of the number of leaves $n$ in trees generated
with stochastic loss events (dots with error bars). Each data point is an average over 100 realizations with error bars indicating standard deviations. For comparison, the expected depth for the ERM model ($\Box$) and for complete binary trees ($\triangle$) are shown.
}
\end{figure}

Suppose the $i$-th innovation, generating feature $i$, affects a species $s $
with $f$ features. Then $s$ is removed from the set $S$ of extant species,
turning into an inner node in the tree. Two new species $s^\prime$ and
$s^{\prime\prime}$ are attached, having feature sets $s^\prime=s$ and
$s^{\prime\prime} = \{ i \} \cup s$. In subsequent loss events, a subtree $T_i$
is built up with $2^f$ leaves, each of which is a species $\sigma \subseteq s
\cup \{i\}$. Call $D(T_i)$ sum of the distances of all the leaves in $T_i$ from
the root of $T_i$.

Let us now estimate the expectation value $\langle D(T_i) \rangle$, which
only depends on the number of features of $f$. Trivially, $D(T_i)$ is lower
bounded by $f 2^f$ since the most compact tree is the fully balanced one with
all nodes at distance $f$ from root. In particular, we conjecture
\begin{equation} \label{eq:subtree_bounds}
f 2^f < \langle D(T_i) \rangle < D_\text{ERM}(2^f)~.
\end{equation}
The second inequality is corroborated by the plots in
Figure~\ref{fig:depth_scaling_loss}. We make it plausible as follows. Similar to
the ERM model, a leaf is chosen in each time step when executing loss events.
Here, however, the loss event is performed only if the chosen leaf carries the
chosen feature and the reduced feature set  is not yet present in the tree.
Thus the probability of accepting a proposed loss event at a leaf $s$ is
anticorrelated with the number of features $|s|$ at $s$. The expected number of
features carried by a leaf decreases with its distance from root. Therefore we
argue that the present model adds new nodes preferentially to leaves closer to
root than average, resulting in trees with an expected depth increasing more
slowly than in the ERM model.

\subsection{Approximation of depth scaling}

\begin{figure}
\centerline{\includegraphics[width=\textwidth]{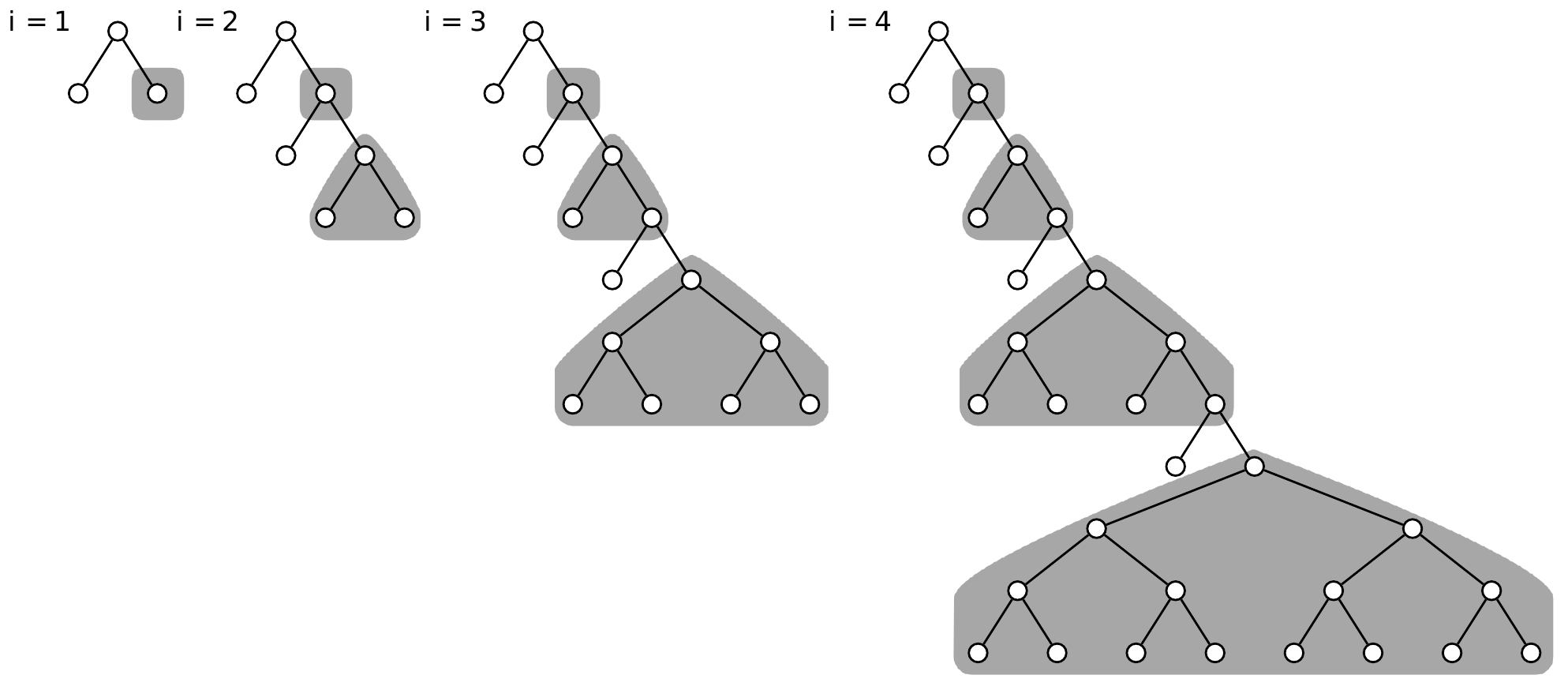}}
\caption{\label{fig:steps_illu}
The deterministic growth of a tree considered as an approximation of
the innovation model. Each subtree generated by an innovation is
indicated as a shaded area.}
\end{figure}

We study a tree growth that is derived from the
innovation model by two simplifying assumptions. (i) Each innovation is
introduced at the leaf with the largest number of features in the tree. (ii) 
Introducing an innovation at a leaf with $f$ features triggers the growth of
a subtree that is a perfect (complete) binary tree with $2^f$ leaves at
distance $f$ from the root of this subtree.

This leads us to consider the following {\em deterministic growth} starting with
a single node and $i=0$. Choose a leaf $s$ at maximum distance from root; split
$s$ obtaining new leaves $s^\prime$ and $s^{\prime\prime}$; take
$s^{\prime\prime}$ as the root of a newly added subtree that is a perfect tree
with $2^i$ leaves; increase $i$ by one and iterate. Figure~\ref{fig:steps_illu}
illustrates the first few steps of the growth.

After $i$ steps, the number of leaves added to the tree most recently 
is $2^{i-1}$. Therefore, the total number of leaves after step $i$ is
\begin{equation}
n(i) = 1 + \sum_{j=1}^i 2^{j-1} = 2^i~.
\end{equation}
because the procedure starts with a single leaf at $i=0$.

The leaves of the subtree added by the $j$-th innovation have distance
\begin{equation}
\sum_{k=1}^j k = \frac{j(j+1)}{2}
\end{equation}
from root because these leaves are $j$ levels deeper than those generated
by the previous innovation. Therefore the sum of all leaves' distances from
root is
\begin{equation} \label{eq:D_of_i}
D(i) = i+ \sum_{j=1}^n 2^{j-1} [j(j+1)/2]
\end{equation}
after the $i$-th innovation has been performed. The first term $i$ arises
because the innovation itself renders one previously existing leaf at a
distance increased by one, cf. the leaves outside the shaded areas in
Figure~\ref{fig:steps_illu}. In performing the sum of
Equation~\ref{eq:D_of_i} we use the equality
\begin{equation}
\sum_{j=0}^i x^{j-1} [j(j+1)] = 2^i [i^2-i+2] -2 
\end{equation}
to arrive at
\begin{equation}
D(i) = i + 2^{i-1} [i^2-i+2] -1~.
\end{equation}
We substitute $n(i)=2^i$, i.e.\ $i=\log_2 n$, and divide $D$ by $n$ to arrive
at the depth
\begin{equation}\label{eq:d_of_n}
d(n) = \frac{1}{2} [ (\log_2 n)^2 - (\log_2 n) +2 ] + \frac{(\log_2 n) -1}{n}  
\end{equation}
% y=sqrt(0.5*(log2(x)^2-log2(x) +2)) for xmgrace
% y=sqrt(0.5*(log2(x)^2-log2(x) +2) + (log2(x)-1) / x) )
of the tree with $n$ leaves generated by deterministic growth. For
large $n$, the depth scaling is
\begin{equation}
d(n) \sim (\log n)^2~.
\end{equation}

\begin{figure}
\centerline{\includegraphics[width=\textwidth]{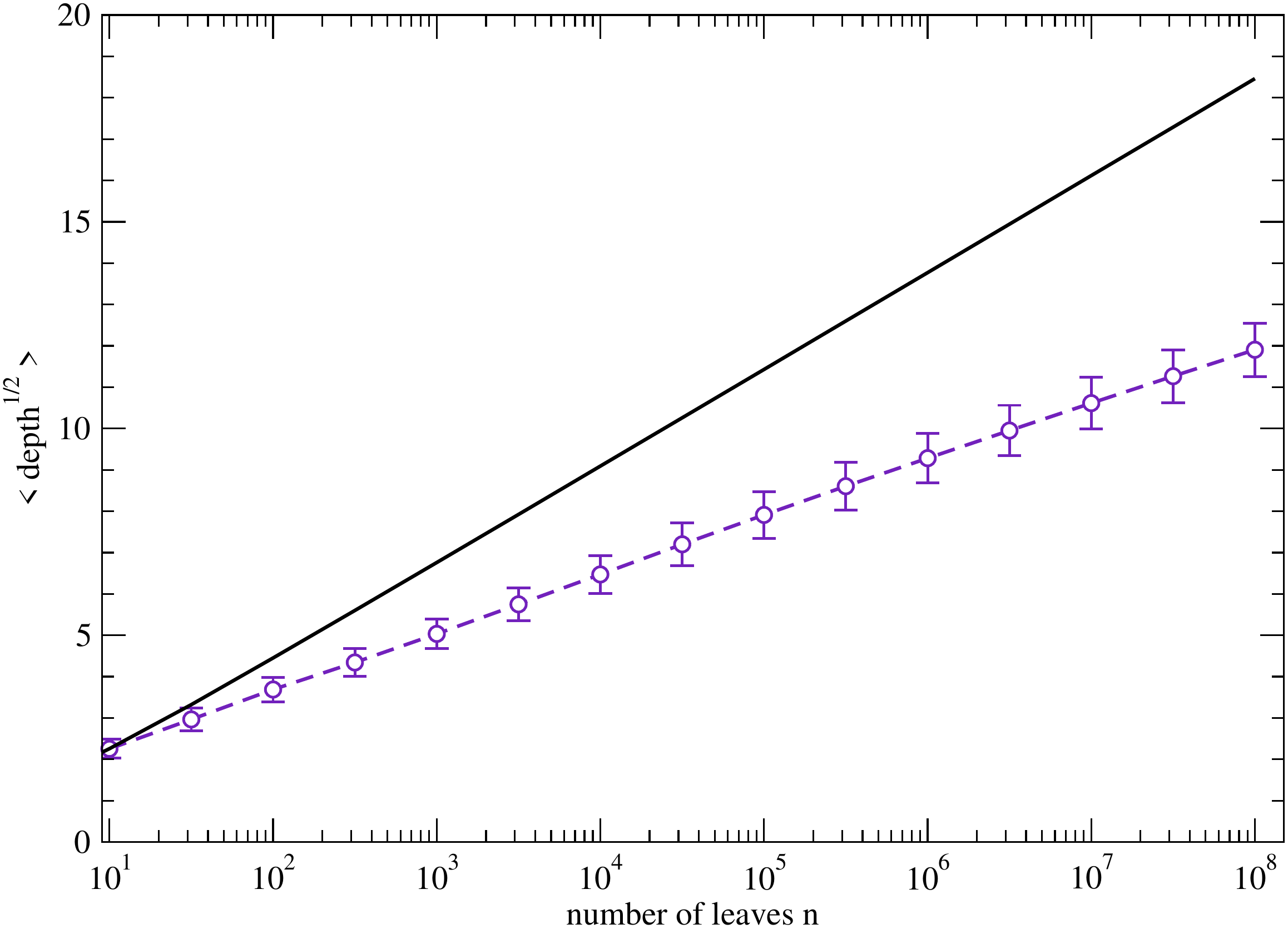}}
\caption{\label{fig:scaling_global}
Depth as a function of tree size $n$ for the innovation model
($\circ$) and for the deterministic growth 
(solid curve) according to Equation~(\ref{eq:d_of_n}). Note
that square root of depth is plotted such that a straight line
in the plot indicates a depth scaling $d(n) \sim (\log n)^2$.
For each size $n$, the plotted point ($\circ$) is the average over
$\sqrt{d(n)}$ for 100 independently generated trees. Error bars
give the standard deviation.
}
\end{figure}

By the comparison in Fig.~\ref{fig:scaling_global}, we find
the $(\log n)^2$ scaling also for the depth of 
trees obtained from the 
innovation model as defined in Section~\ref{sec:innov_def}.
Thus we hypothesize that the deterministic growth captures the
essential mechanism leading to the depth scaling of the innovation model.
The prefactor of $(\log n)^2$ is smaller in the innovation model than
in the deterministic growth. In the actual model, most innovations hit a
leaf with a non-maximal number of features and therefore trigger the
growth of a lower subtree than assumed by deterministic growth.
Table~\ref{tab:depth} provides an overview of the scaling of average depth with
the number of leaves for various tree models .

\begin{table} 
\renewcommand{\arraystretch}{1.5}
\caption{\label{tab:depth} Depth scaling of models.} 
\begin{tabular}{@{}ll@{}} 
\hline
\bf{innovation model}	        & $(\log n)^2$  \\
$\beta$-\bf{splitting} \cite{ALDOUSBetaSplitting1996}& $\begin{cases} 
\log n        & \text{if } \beta > -1 \text{, includes \bf{ ERM} }(\beta=0)\\
(\log n)^2    & \text{if } \beta = -1 \text{, \bf{AB} model}\\
n^{-\beta-1}  & \text{if } \beta < -1 \text{, includes \bf{ PDA} }(\beta=-1.5)
\end{cases}$\\
\bf{age model} \cite{KellerSchmidtTugrul2010AgeModel}		& $(\log n)^2$ 	\\
\bf{activity model} \cite{HernandezGarcia2010scaling}
& $\begin{cases} n^{0.5} & \text{if } p=0.5,\\ \log n  & \text{otherwise.} \end{cases}$ \\
\bf{complete tree}              & $\log n$\\
\bf{comb tree}                  & $n$\\
\hline
\end{tabular}
\end{table}

%%%%%%%%%%%%%%%%%%%%%%%%%%%%%%%%%%%%%%%%%%%%%%%%%%%%%%%%%%%%%%%%%%%%%%%%%%%%
\section{Discussion}
%%%%%%%%%%%%%%%%%%%%%%%%%%%%%%%%%%%%%%%%%%%%%%%%%%%%%%%%%%%%%%%%%%%%%%%%%%%%

The innovation model establishes a connection between the burstiness of
macroevolution and the observed imbalance of phylogenetic trees. Bursts of
diversification are triggered by generation of new features and combination
with the repertoire of existing traits. In order to keep the model simple,
the diversification after an innovation is implemented as a sequence of
random losses of features. More realistic versions of the model could be
studied where combinations of traits are enriched by re-activation of
previously silenced traits or horizontal transfer between species.
Furthermore, the model as presented here neglects the extinction of species
and their influence on the shapes of phylogenetic trees.

Regarding the robustness of the model, the depth scaling would have to be
tested under modifications. In particular, the infinite time
scale separation between rare innonvations and frequent loss events
could be given up by allowing innovations to occur at a finite rate
set as a parameter.

In summary, we have defined a well-working, biologically motivated model which
nevertheless is sufficiently simple to allow for further enhancement
regarding biological concepts such as sequence evolution and
genotype-phenotype relations.

\section*{Acknowledgments}

The authors thank Kathrin Lembcke, Maribel Hern\'{a}ndez Rosales and Nicolas
Wieseke for a critical reading of the draft. This work was supported by
Volkswagen Stiftung through the initiative for Complex Networks as a Phenomenon
across Disciplines.

\bibliography{innovmodelPaper}
\bibliographystyle{abbrv}

\end{document}